\documentclass[conference]{IEEEtran}
\IEEEoverridecommandlockouts

\usepackage{cite}
\usepackage{amsmath,amssymb,amsfonts}
\usepackage{algorithmic}
\usepackage{graphicx}
\usepackage{textcomp}
\usepackage{xcolor}
\usepackage{booktabs}
\usepackage{multirow}

\def\BibTeX{{\rm B\kern-.05em{\sc i\kern-.025em b}\kern-.08em
    T\kern-.1667em\lower.7ex\hbox{E}\kern-.125emX}}
\begin{document}

\title{Multi-view Hypergraph-based Contrastive Learning Model for Cold-Start Micro-video Recommendation\\
{}
\thanks{This work was supported by the Guangdong Provincial Department of Education Project (Grant No.2024KQNCX028).}
\thanks{* Corresponding author}

}

\author{\IEEEauthorblockN{Sisuo LYU}
\IEEEauthorblockA{\textit{The Hong Kong University of}\\
\textit{Science and Technology (Guangzhou)}\\
Guangzhou, China \\
sisuolyu@outlook.com}
\and
\IEEEauthorblockN{Xiuze ZHOU}
\IEEEauthorblockA{\textit{The Hong Kong University of}\\
\textit{Science and Technology (Guangzhou)}\\
Guangzhou, China \\
zhouxiuze@foxmail.com}
\and
\IEEEauthorblockN{Xuming HU*}
\IEEEauthorblockA{\textit{The Hong Kong University of}\\
\textit{Science and Technology (Guangzhou)}\\
Guangzhou, China \\
xuminghu@hkust-gz.edu.cn}

}

\maketitle

\begin{abstract}
With the widespread use of mobile devices and the rapid growth of micro-video platforms such as TikTok and Kwai, the demand for personalized micro-video recommendation systems has significantly increased. Micro-videos typically contain diverse information, such as textual metadata, visual cues (e.g., cover images), and dynamic video content, significantly affecting user interaction and engagement patterns. However, most existing approaches often suffer from the problem of over-smoothing, which limits their ability to capture comprehensive interaction information effectively. Additionally, cold-start scenarios present ongoing challenges due to sparse interaction data and the underutilization of available interaction signals.

To address these issues, we propose a Multi-view Hypergraph-based Contrastive learning model for cold-start micro-video Recommendation (MHCR). MHCR introduces a multi-view multimodal feature extraction layer to capture interaction signals from various perspectives and incorporates multi-view self-supervised learning tasks to provide additional supervisory signals. Through extensive experiments on two real-world datasets, we show that MHCR significantly outperforms existing video recommendation models and effectively mitigates cold-start challenges. Our code is available at https://github.com/sisuolv/MHCR.

\end{abstract}


\section{Introduction}
In recent years, the rapid expansion of platforms such as TikTok and Kwai has heightened the need for effective personalized micro-video recommendation systems. Unlike other content types, such as news or music, micro-videos involve richer multi-modal features, typically including titles (text), cover images (visual), and the videos themselves (visual). These multi-modal features are crucial in shaping users' behavioral decisions and largely determine whether a user engages with a micro-video \cite{yi2022multi, li2022improving, patil2024micro,ge2024mambatsr,zhou2023mmrec}.

As a specialized area of recommendation systems, video recommendation has made numerous valuable advancements. For instance, MMGCN \cite{wei2019mmgcn} integrates multi-modal features into a user--item bipartite graph neural network to refine user--item interactions, mitigating the impact of spurious interactions on model performance. MMGCL \cite{yi2022multi} introduces a novel negative sampling technique to learn cross-modal correlations, ensuring each modality contributes effectively and enhancing recommendation accuracy. CMI \cite{li2022improving} leverages contrastive learning across multiple interests to capture users' diverse preferences, improving the robustness of interest embeddings.

Despite the progress made by existing micro-video recommendation methods, they still face significant challenges in cold-start scenarios, primarily due to two issues:

\begin{itemize}
\item \textbf{Sparse interaction signals}: Micro-video recommendation is heavily impacted by the long-tail distribution, where most micro-videos receive minimal interaction signals. This results in suboptimal performance of current models in cold-start conditions.
\item \textbf{Underutilization of interaction information}: Most existing methods rely on Graph Neural Networks (GNNs) to aggregate interaction information through message passing. However, these methods often suffer from over-smoothing, limiting the model's ability to capture comprehensive interaction information.
\end{itemize}

To address these challenges, we propose a \textbf{M}ulti-view \textbf{H}ypergraph-based \textbf{C}ontrastive Learning Model for Cold-Start Micro-video \textbf{R}ecommendation (MHCR). First, MHCR employs hypergraphs to capture more extensive interaction information, facilitating the effective propagation of interaction signals. Second, MHCR incorporates multiple self-supervised learning tasks to provide additional supervision signals, enhancing the model's learning capability. Our key contributions are as follows:

\begin{itemize}
\item We introduce MHCR, the first model to leverage hypergraphs and contrastive learning to tackle the cold-start problem in micro-video recommendation.
\item We design a multi-view multimodal feature extraction layer to repeatedly harness interaction information from multiple perspectives and introduce multi-view self-supervised learning tasks to provide auxiliary supervision signals.
\item Extensive experiments on two real-world datasets validate the effectiveness of MHCR and ablation studies confirm the contribution of each module within the model.
\end{itemize}

\begin{figure*}[t]
    \centering
    \includegraphics[height=0.3\textheight, width=0.8\textwidth]{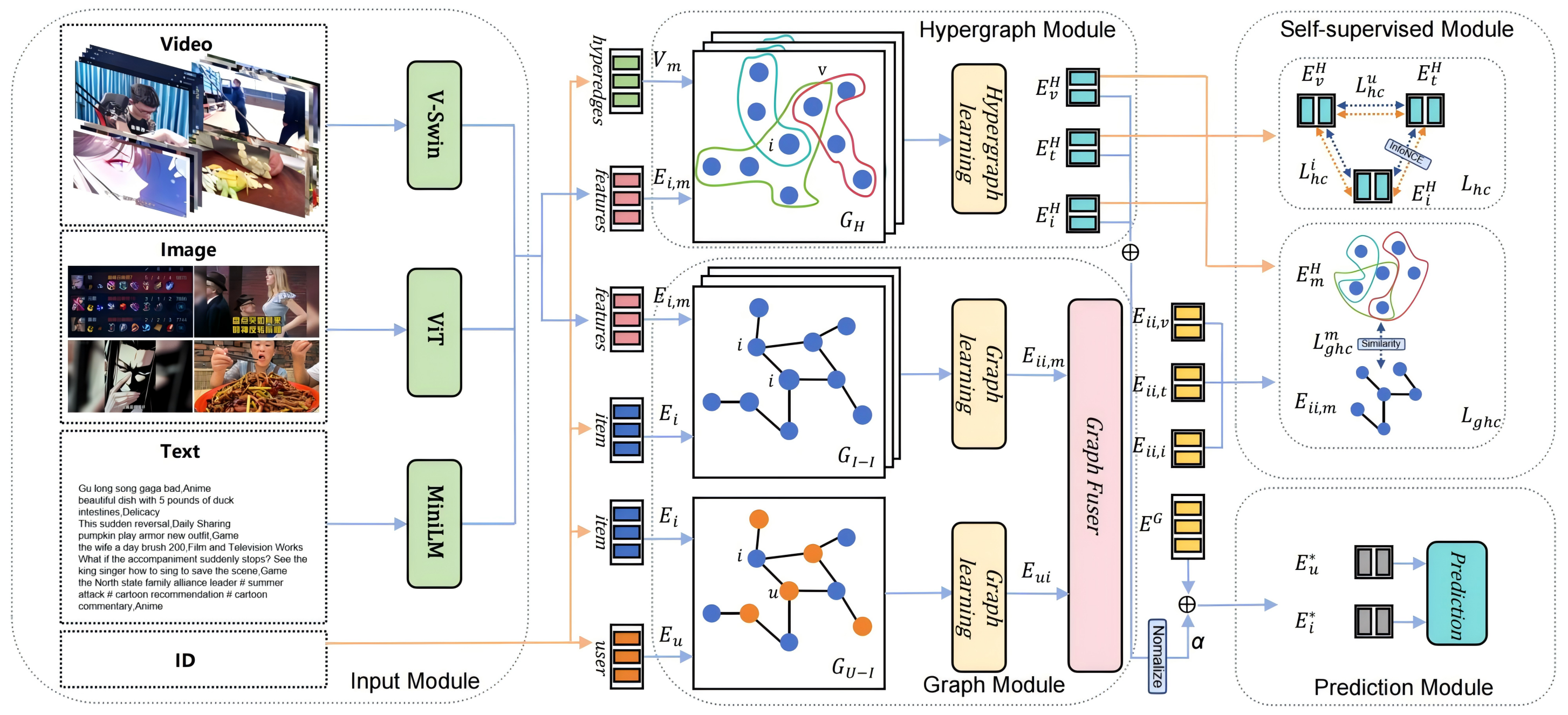} 
    \caption{Overview of the MHCR Architecture.}
    \label{fig:overview}
\end{figure*}
\section{PROPOSED METHOD}

\subsection{Problem Definition}
Let \( U = \{u\} \) and \( I = \{i\} \) represent the sets of users and items, respectively. The corresponding ID embeddings are represented as \( E_{ui} \in \mathbb{R}^{d \times (|U| + |I|)} \), where \( d \) denotes the embedding dimension. Additionally, we define \( M = \{i, v, t\} \), which encompasses the image, video, and text modalities. For each modality \( m \in M \), the item modality features are represented as \( E_{i,m} \in \mathbb{R}^{d_m \times |I|} \). These features are projected into a unified vector space by pre-trained models (e.g., ViT for images, V-Swin for videos, and MiniLM for text), and multimodal features are extracted from multi-view. The framework is shown in Fig.~\ref{fig:overview}.

\subsection{Multi-view Multimodal Feature Extraction Layer}
\label{ssec:MIE}

\noindent \textbf{User--Item Graph Feature Extraction Layer.} Inspired by LightGCN, we design a user--item graph aimed at effectively capturing high-order collaborative signals, and the representation at the \(l\)-th layer is defined as follows:

\begin{equation}
    E_{ui}^{(l)} = \sum_{i \in N_u} \frac{1}{\sqrt{|N_u| \cdot |N_i|}} E_{ui}^{(l-1)},
\end{equation}
where \(N_u\) and \(N_i\) denote the neighbors of user \(u\) and item \(i\), respectively. 

The final user--item representation is obtained by summing the embeddings from all layers: \(E_{ui} = \sum_{l=0}^{L} E_{ui}^{(l)}\), where \(L\) is the total number of graph convolution layers.

\noindent \textbf{Item--Item Graph Feature Extraction Layer.} Similar to the user--item view, we construct an item--item affinity graph to capture modality-specific correlations, and the affinity between items \(a\) and \(b\) in modality \(m\) is defined as follows:

\begin{equation}
    s_{a,b}^m = \frac{(e_a^m)^\text{T} \cdot e_b^m}{\|e_a^m\| \|e_b^m\|},
\end{equation}
where \(e_a^m\) and \(e_b^m\) are the modality features of items \(a\) and \(b\). 

We apply KNN sparsification to keep the top \(K\) nearest neighbors for each item and normalize the affinity matrix. The item modality features (\(E_{i,m}\)) are then propagated through the graph. The overall item embedding is computed as:
$E_{ii} = \sum_m S_m E_{i,m}$.

\begin{table*}[t]
\centering
\caption{Performance Metrics of Different Models on Two MicroLens Datasets}
\setlength{\tabcolsep}{3pt} 
\begin{tabular}{llcccccccccc}
\toprule
\textbf{Dataset} & \textbf{Metric} & \textbf{YouTube} & \textbf{VBPR} & \textbf{MMGCN} & \textbf{LightGCN} & \textbf{GRCN} & \textbf{LayerGCN} & \textbf{BM3} & \textbf{Freedom} & \textbf{MGCN} & \textbf{MHCR} \\
\midrule
\multirow{4}{*}{\textbf{MicroLens-50K}} 
                                    & R@10  & 0.0375 & 0.0544 & 0.0403 & 0.0365 & 0.0631 & 0.0627 & 0.0565 & 0.0656 & 0.0708 & \textbf{0.0736} \\
                                    & R@20  & 0.0632 & 0.0888 & 0.067 & 0.0534 & 0.0982 & 0.0994 & 0.0918 & 0.1028 & 0.1089 & \textbf{0.1102} \\
                                    & N@10    & 0.0178 & 0.0273 & 0.0197 & 0.0284 & 0.0328 & 0.032 & 0.0281 & 0.0334 & 0.0363 & \textbf{0.0383} \\
                                    & N@20    & 0.0245 & 0.0361 & 0.0264 & 0.0345 & 0.0415 & 0.0414 & 0.0372 & 0.0429 & 0.0459 & \textbf{0.0477} \\
\midrule
\multirow{4}{*}{\textbf{MicroLens-100K}} 
                                    & R@10  & 0.0392 & 0.0624 & 0.0405 & 0.0388 & 0.0682 & 0.0730 & 0.0601 & 0.0654 & 0.0717 & \textbf{0.0798} \\
                                    & R@20  & 0.0648 & 0.1002 & 0.0678 & 0.056 & 0.1057 & 0.1120 & 0.0975 & 0.1016 & 0.1096 & \textbf{0.1187} \\
                                    & N@10    & 0.0188 & 0.0314 & 0.0202 & 0.0306 & 0.0353 & 0.0382 & 0.0305 & 0.0337 & 0.0371 & \textbf{0.042} \\
                                    & N@20    & 0.0252 & 0.0410 & 0.0271 & 0.0367 & 0.0448 & 0.0480 & 0.0401 & 0.0431 & 0.0467 & \textbf{0.0519} \\

\bottomrule
\end{tabular}
\label{tab:model_metrics}
\end{table*}

\noindent \textbf{Hypergraph High-Level Feature Extraction Layer.} To capture higher-order dependencies between users, items, and their attributes, we introduce learnable hyperedge embeddings: \( H_i^m = E_i^m \cdot V_m^\top \), \(H_u^m = X_u \cdot (H_i^m)^\top\), where \(E_i^m\) is the item feature matrix for modality \(m\), and \(V_m \in \mathbb{R}^{K \times d_m}\) represents the hyperedge vector matrix, with \(K\) denoting the number of hyperedges. Additionally, \(X_u\) is derived from the user-item interaction matrix \(X\). Hypergraph message passing facilitates global information transfer by using hyperedges:

\begin{equation}
\begin{cases}
    E_i^{m,h+1} = DROP(H_i^m) \cdot DROP((H_i^m)^\top) \cdot E_i^{m,h}, \\
    E_u^{m,h+1} = DROP(H_u^m) \cdot DROP((H_i^m)^\top) \cdot E_i^{m,h},
\end{cases}
\end{equation}

The hypergraph embedding matrix \(E_{h}\) is obtained by aggregating the embeddings across all modalities:
\\$E_{h} = \sum_{m } E_m^H, \quad E_m^H = [E_u^m, E_i^m].$

By concatenating the user--item features \(E_{ui}\), item--item features \(E_{ii}\), and hypergraph embedding \(E_{h}\), the final modality feature \(E^* \in \mathbb{R}^{d \times (|U| + |I|)}\) is obtained.

\subsection{Multi-view Self-Supervised Learning}
\label{ssec:subhead}

To ensure the effective fusion of global embeddings across various modalities, we introduce a cross-modal hypergraph contrastive learning:

\begin{equation}
    L_{hc} = \sum_{x \in U \cup I} -\log \frac{\sum_m\exp\left(\frac{s(E_x^{m,H}, E_x^{m',H})}{\tau}\right)}{\sum_{x' \in U \cup I} \sum_m\exp\left(\frac{s(E_x^{m,H}, E_x^{t,H})}{\tau}\right)},
\end{equation}
where \( s(\cdot) \) denotes the cosine similarity; \( \exp \) denotes the exponential function; and \( \tau \) denotes the temperature parameter.

To align graph and hypergraph embeddings, a graph-hypergraph contrastive learning strategy is used, and the loss is defined as follows:

\begin{align}
L_{ghc} = \sum \left( -\log \frac{\exp\left( \frac{s(E^{G}, E^{H})}{\tau} \right)}{\exp\left( \frac{s(E^{G}, E^{H})}{\tau} \right) + \sum \exp\left( \frac{s(E^{G}, E^{H})}{\tau} \right)} \right).
\end{align}

Finally, the overall loss combines the BPR loss \(L_{BPR}\) with the hypergraph contrastive loss \(\lambda_{hc} L_{hc}\) and the graph hypergraph contrastive loss \(\lambda_{ghc} L_{ghc}\), ensuring a balance between ranking and contrastive learning.

\subsection{Prediction}
\label{ssec:prediction}

Based on the refined behavioral and multimodal features, the final representations for users and items are formulated as follows: \(e_u = e_{u,ui} + e_{u,ii}+ e_{u, h}, \quad e_i = e_{i, ui} + e_{i,ii} + e_{i, h}\). The interaction likelihood between user \(u\) and item \(i\) is determined by computing their inner product: 
\begin{equation}
f_{predict}(u, i) = \hat{y}_{ui} = e_u^\top e_i.
\end{equation}

\section{EXPERIMENTS}
In this section, we conduct experiments to assess the effectiveness of the proposed model across two MicroLens datasets. The experimental results provide clear insights into the following research questions: (1) \textbf{RQ1:} How does the MHCR's performance compare to existing video recommendation techniques?
(2) \textbf{RQ2:} What is the impact of the individual components on the overall performance of the MHCR?
(3) \textbf{RQ3:} How do variations in hyper-parameter configurations affect the MHCR's outcomes?
(4) \textbf{RQ4:} How does the performance of the MHCR model in cold-start scenarios compare to baseline models?

\subsection{Experimental Settings}
\subsubsection{Datasets}
In this study, we utilized two datasets from the MicroLens series, namely MicroLens-50K and MicroLens-100K. The MicroLens-50K dataset comprises 50,000 users, 19,220 items, and a total of 359,708 interaction records, resulting in an average of 7.19 interactions per user and 18.71 interactions per item, with an overall sparsity level of 99.96\%. Similarly, the MicroLens-100K dataset includes 100,000 users, 19,738 items, and 719,405 interaction records. Beyond the interaction data, the MicroLens datasets also feature 15,580 tags representing detailed video categories. The length of interaction sequences typically falls between 5 and 15, with most micro-videos lasting less than 400 seconds, providing a more comprehensive contextual dataset for in-depth analysis.

\subsubsection{Compared methods}
To verify the effectiveness of the proposed MHCR, we conducted a detailed comparison of three different types of recommendation models: (1) {General Deep Learning Recommendation Models}, including \textbf{YouTube}\cite{covington2016deep} and \textbf{VBPR}\cite{he2016vbpr}; (2) {Graph-based Recommendation Models}, including \textbf{LightGCN}\cite{he2020lightgcn} and \textbf{LayerGCN}\cite{zhou2023layer}; and (3) {Multimodal Graph-based Recommendation Models}, including \textbf{MMGCN}\cite{wei2019mmgcn}, \textbf{GRCN}\cite{wei2020graph}, \textbf{BM3}\cite{zhou2023bootstrap}, \textbf{Freedom}\cite{zhou2023tale}, and \textbf{MGCN}\cite{yu2023multi}.

\subsubsection{Evaluation Protocols}

To evaluate model performance, we randomly split each user's interaction history into 70\% for training, 10\% for validation, and 20\% for testing. Recall@\(n\) and NDCG@\(n\) (with \(n \in \{10, 20\}\)) are used as evaluation metrics to assess the effectiveness of the top-\(n\) recommendations. In the following discussions, we use R@\(n\) and N@\(n\) as shorthand for Recall@\(n\) and NDCG@\(n\), respectively.

\subsection{Overall Performance (RQ1)}

Based on the results presented in Table~\ref{tab:model_metrics}, we compare the performance of the proposed MHCR model with baseline models and highlight the following key findings:

\begin{enumerate}

\item The MHCR model consistently outperforms all baseline models across the MicroLens-50K and MicroLens-100K datasets. Specifically, on the MicroLens-50K dataset, MHCR demonstrates a significant improvement of 3.96\% in Recall@10 and 5.51\% in NDCG@10 when compared to the closest baseline model, MGCN. Similarly, on the larger MicroLens-100K dataset, MHCR outperforms other models with an increase of 11.30\% in Recall@10 and 13.19\% in NDCG@10. These results highlight the effectiveness of the proposed Hypergraph Embedding Module and the Self-Supervised Learning mechanism, which allow MHCR to capture better the sparse interaction patterns inherent in micro-video recommendation tasks.

\item Multimodal Graph-based Recommendation Models generally outperform Graph-based Recommendation Models, underscoring the importance of incorporating multimodal information, which plays a pivotal role in enhancing the predictive performance of recommendation systems.

\item Graph-based Recommendation Models consistently surpass general deep learning recommendation models, highlighting the capability of graph neural networks to effectively aggregate neighborhood information, thus enabling more precise modeling of user and item representations.
\end{enumerate}

\subsection{Ablation Study(RQ2)}

To rigorously evaluate the contribution of each component in our proposed MHCR model, we conducted an ablation study by systematically disabling each module and analyzing the impact on performance. The experiments were conducted on the MicroLens-100K dataset using the following configurations:
(1) \textbf{w/o UI}: removing the User--Item Graph Embedding Module. 
(2) \textbf{w/o II}: removing the Item--Item Graph Embedding Module.
(3) \textbf{w/o HEM}: removing the Hypergraph Embedding Module.
(4) \textbf{w/o HC}: removing the Hypergraph Contrastive Module.
(5) \textbf{w/o GHC}: removing the Graph-Hypergraph Contrastive Module.

\begin{figure}[ht!]
    \centering
    \includegraphics[width=1\linewidth]{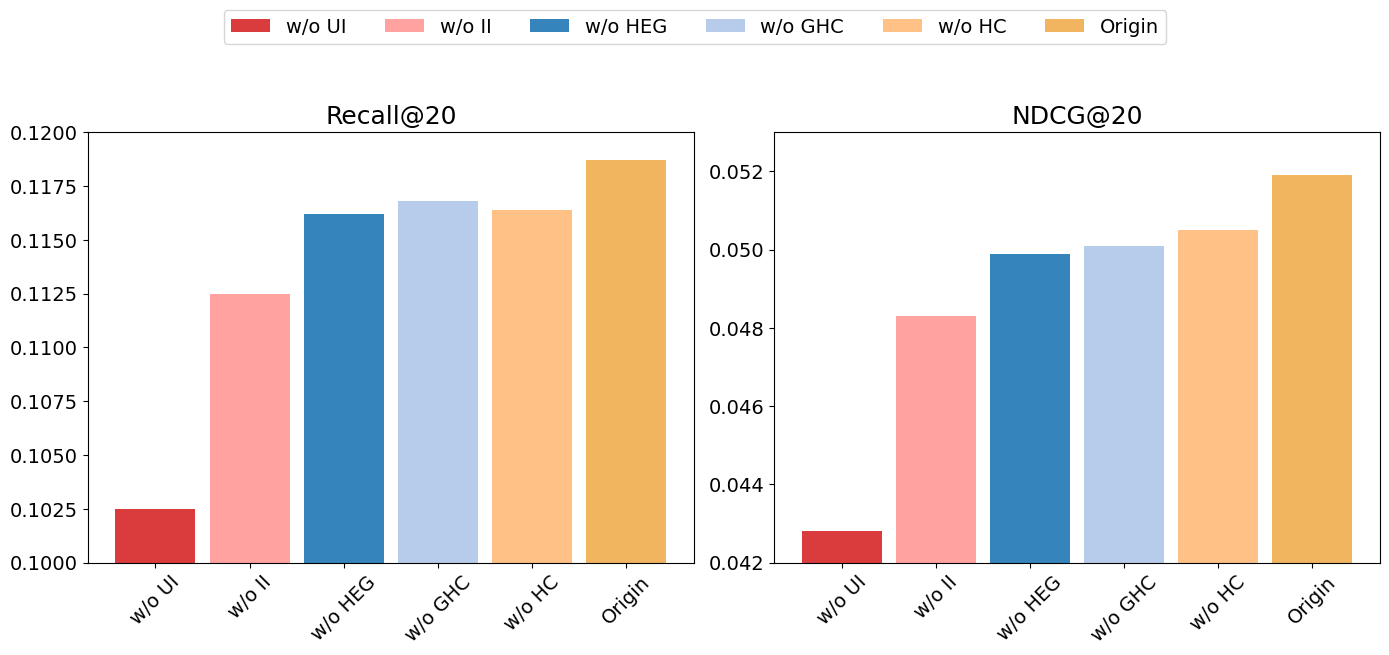}
    \caption{Ablation Study Results of MHCR.}
    \label{fig:metrics_comparison}
\end{figure}

Fig. \ref{fig:metrics_comparison} illustrates the results of our ablation study on the MicroLens-100K. The original MHCR configuration consistently outperforms its modified versions, underscoring the critical importance of each component. The performance drop observed when individual modules are removed respectively highlights the key roles of the graph, hypergraph, and self-supervised learning components, confirming their important contributions to the overall effectiveness of the MHCR model.

\subsection{Sensitivity Analysis (RQ3)}

In this section, we perform a series of experiments to identify the optimal hyperparameters, explicitly focusing on the number of hypergraph layers (\(\boldsymbol{hyper\_num} \)), the ratio of the hypergraph contrastive loss (\( \boldsymbol{\lambda_{hc}}\)), and the ratio of the graph-hypergraph contrastive loss (\(\boldsymbol{ \lambda_{ghc}} \)).

\begin{figure}[ht!]
    \centering
    \includegraphics[width=1.05\linewidth]{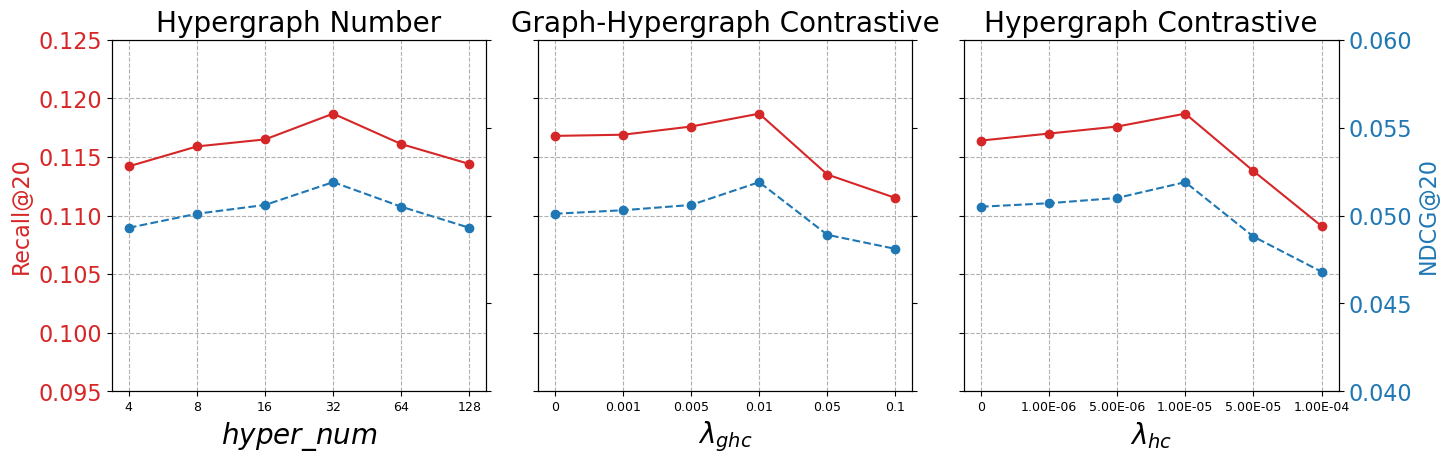}
    \caption{Effect of Hypergraph Parameters and Contrastive Learning on Recommendation Performance.}
    \label{fig:Parameter}
\end{figure}

The results presented in Fig.\ref{fig:Parameter} reveal three key insights: First, setting \(hyper\_num\) to 32 yields the best performance while increasing the number of layers slightly degrades performance due to increased complexity. Second, the optimal value for \( \lambda_{hc}\) is found to be \( 1.00 \times 10^{-5} \), which enhances the effectiveness of contrastive learning without leading to overfitting. Finally, \( \lambda_{ghc}\) performs optimally at 0.01, effectively aligning representations, while higher values result in diminished performance.

\subsection{Results on Cold-start Scenarios (RQ4)}
In the cold-start scenario, where users have fewer than three interactions, the experimental results across the two datasets demonstrate that the MHCR model consistently outperforms the baselines, as shown in Table~\ref{tab:cold-start}. On the MicroLens-50K dataset, MHCR improves Recall@10 and NDCG@10 by 4.8\% and 7.3\%, respectively. Similarly, on the MicroLens-100K dataset, MHCR improves by 5.0\% in Recall@20 and 6.8\% in NDCG@20. These results highlight MHCR's effectiveness in addressing cold-start challenges through its multi-view hypergraph contrastive learning approach.

\begin{table}[!t]
\centering
\caption{Performance Comparison of Different Models for Cold-Start Users on Two MicroLens Datasets}
\setlength{\tabcolsep}{3pt} 
\begin{tabular}{llcccccccc}
\toprule
\textbf{Dataset} & \textbf{Metric} & \textbf{GRCN} & \textbf{LayerGCN} & \textbf{BM3} & \textbf{Freedom} & \textbf{MGCN} & \textbf{MHCR} \\
\midrule
\multirow{4}{*}{\textbf{M-50K}} 
                                    & R@10  & 0.0521 & 0.0527 & 0.0471 & 0.053 & 0.0588 & \textbf{0.0616} \\
                                    & R@20  & 0.081 & 0.0826 & 0.0789 & 0.0839 & 0.0912 & \textbf{0.0937} \\
                                    & N@10    & 0.0267 & 0.0267 & 0.0237 & 0.0267 & 0.0299 & \textbf{0.0321} \\
                                    & N@20    & 0.034 & 0.0341 & 0.0316 & 0.0344 & 0.038 & \textbf{0.0396} \\
\midrule
\multirow{4}{*}{\textbf{M-100K}} 
                                    & R@10  & 0.0562 & 0.0588 & 0.0485 & 0.0544 & 0.0605 & \textbf{0.0655} \\
                                    & R@20  & 0.0874 & 0.092 & 0.0807 & 0.0882 & 0.0937 & \textbf{0.0984} \\
                                    & N@10    & 0.0288 & 0.0299 & 0.024 & 0.0275 & 0.0311 & \textbf{0.0342} \\
                                    & N@20    & 0.0366 & 0.0384 & 0.0321 & 0.0359 & 0.0396 & \textbf{0.0423} \\
\bottomrule
\end{tabular}
\label{tab:cold-start}
\end{table}

\section{Conclusion}

To address the cold-start issue and limitations in capturing comprehensive interaction information, this paper presents MHCR, which leverages hypergraph structures for intricate interaction patterns and self-supervised learning to improve predictive performance. It enhances embeddings through user--item and item--item graphs for multi-hop connections and intra-modal relationships. Hypergraph embeddings capture higher-order associations, while contrastive losses boost robustness and distinctiveness across modalities. Experiments on real-world datasets confirm MHCR's effectiveness in handling multimodal data and cold-start challenges, paving the way for further improvements in hypergraph-based recommendation systems.


\newpage
\bibliographystyle{IEEEtran} 
\bibliography{refs} 

\begin{thebibliography}{10}
\providecommand{\url}[1]{#1}
\csname url@samestyle\endcsname
\providecommand{\newblock}{\relax}
\providecommand{\bibinfo}[2]{#2}
\providecommand{\BIBentrySTDinterwordspacing}{\spaceskip=0pt\relax}
\providecommand{\BIBentryALTinterwordstretchfactor}{4}
\providecommand{\BIBentryALTinterwordspacing}{\spaceskip=\fontdimen2\font plus
\BIBentryALTinterwordstretchfactor\fontdimen3\font minus \fontdimen4\font\relax}
\providecommand{\BIBforeignlanguage}[2]{{%
\expandafter\ifx\csname l@#1\endcsname\relax
\typeout{** WARNING: IEEEtran.bst: No hyphenation pattern has been}%
\typeout{** loaded for the language `#1'. Using the pattern for}%
\typeout{** the default language instead.}%
\else
\language=\csname l@#1\endcsname
\fi
#2}}
\providecommand{\BIBdecl}{\relax}
\BIBdecl

\bibitem{yi2022multi}
Z.~Yi, X.~Wang, I.~Ounis, and C.~Macdonald, ``Multi-modal graph contrastive learning for micro-video recommendation,'' in \emph{Proceedings of the 45th International ACM SIGIR Conference on Research and Development in Information Retrieval}, 2022, pp. 1807--1811.

\bibitem{li2022improving}
B.~Li, B.~Jin, J.~Song, Y.~Yu, Y.~Zheng, and W.~Zhou, ``Improving micro-video recommendation via contrastive multiple interests,'' in \emph{Proceedings of the 45th International ACM SIGIR Conference on Research and Development in Information Retrieval}, 2022, pp. 2377--2381.

\bibitem{patil2024micro}
S.~S. Patil, R.~S. Patil, and A.~Kotwal, ``Micro video recommendation in multimodality using dual-perception and gated recurrent graph neural network,'' \emph{Multimedia Tools and Applications}, vol.~83, no.~17, pp. 51\,559--51\,588, 2024.

\bibitem{ge2024mambatsr}
Y.~Ge, Z.~Chen, M.~Yu, Q.~Yue, R.~You, and L.~Zhu, ``Mambatsr: You only need 90k parameters for traffic sign recognition,'' \emph{Neurocomputing}, p. 128104, 2024.

\bibitem{zhou2023mmrec}
X.~Zhou, ``Mmrec: Simplifying multimodal recommendation,'' in \emph{Proceedings of the 5th ACM International Conference on Multimedia in Asia Workshops}, 2023, pp. 1--2.

\bibitem{wei2019mmgcn}
Y.~Wei, X.~Wang, L.~Nie, X.~He, R.~Hong, and T.-S. Chua, ``Mmgcn: Multi-modal graph convolution network for personalized recommendation of micro-video,'' in \emph{Proceedings of the 27th ACM international conference on multimedia}, 2019, pp. 1437--1445.

\bibitem{covington2016deep}
P.~Covington, J.~Adams, and E.~Sargin, ``Deep neural networks for youtube recommendations,'' in \emph{Proceedings of the 10th ACM conference on recommender systems}, 2016, pp. 191--198.

\bibitem{he2016vbpr}
R.~He and J.~McAuley, ``Vbpr: visual bayesian personalized ranking from implicit feedback,'' in \emph{Proceedings of the AAAI conference on artificial intelligence}, 2016.

\bibitem{he2020lightgcn}
X.~He, K.~Deng, X.~Wang, Y.~Li, Y.~Zhang, and M.~Wang, ``Lightgcn: Simplifying and powering graph convolution network for recommendation,'' in \emph{Proceedings of the 43rd International ACM SIGIR conference on research and development in Information Retrieval}, 2020, pp. 639--648.

\bibitem{zhou2023layer}
X.~Zhou, D.~Lin, Y.~Liu, and C.~Miao, ``Layer-refined graph convolutional networks for recommendation,'' in \emph{2023 IEEE 39th International Conference on Data Engineering (ICDE)}.\hskip 1em plus 0.5em minus 0.4em\relax IEEE, 2023, pp. 1247--1259.

\bibitem{wei2020graph}
Y.~Wei, X.~Wang, L.~Nie, X.~He, and T.-S. Chua, ``Graph-refined convolutional network for multimedia recommendation with implicit feedback,'' in \emph{Proceedings of the 28th ACM international conference on multimedia}, 2020, pp. 3541--3549.

\bibitem{zhou2023bootstrap}
X.~Zhou, H.~Zhou, Y.~Liu, Z.~Zeng, C.~Miao, P.~Wang, Y.~You, and F.~Jiang, ``Bootstrap latent representations for multi-modal recommendation,'' in \emph{Proceedings of the ACM Web Conference 2023}, 2023, pp. 845--854.

\bibitem{zhou2023tale}
X.~Zhou and Z.~Shen, ``A tale of two graphs: Freezing and denoising graph structures for multimodal recommendation,'' in \emph{Proceedings of the 31st ACM International Conference on Multimedia}, 2023, pp. 935--943.

\bibitem{yu2023multi}
P.~Yu, Z.~Tan, G.~Lu, and B.-K. Bao, ``Multi-view graph convolutional network for multimedia recommendation,'' in \emph{Proceedings of the 31st ACM International Conference on Multimedia}, 2023, pp. 6576--6585.

\end{thebibliography}

\vspace{12pt}

\end{document}